# Melting Point and Lattice Parameter Shifts in Supported Metal Nanoclusters

*V. D. Borman, I. V. Tronin, V. N. Tronin, V. I. Troyan, O. S. Vasiliev, P. V. Borisyuk, M. A. Pushkin*
*(Moscow Engineering Physics Institute)*

The dependencies of the melting point and the lattice parameter of supported metal nanoclusters as functions of clusters height are theoretically investigated in the framework of the uniform approach. The vacancy mechanism describing the melting point and the lattice parameter shifts in nanoclusters with decrease of their size is proposed. It is shown that under the high vacuum conditions ($p < 10^{-7}$ torr) the essential role in clusters melting point and lattice parameter shifts is played by van der Waals forces of cluster-substrate interaction. The proposed model satisfactorily accounts for the experimental data.

The investigation of the properties of supported metal nanoclusters attracts now the increased attention, first of all motivated by the need in the fundamental physical basis for the further development of nanotechnology related to single-electron and nanoelectronic devices, new electronic and magnetic nanomaterials, nanocluster catalysts and nanofilms.

One of the interesting properties of nanometer-sized systems is a shift of the melting point and the lattice parameter with decrease of cluster size (size effect). This effect was experimentally observed for Au, Ag and Cu clusters deposited onto various substrates (W, C) [1-6]. The existing theoretical descriptions of the observed size effect are based on various approaches [1, 2, 7-11]. However, their common and essential drawback is that the influence of the substrate is not taken into account. At the same time it is experimentally proved [6] that substrate plays an important role and the use of different substrates (W, C) results in the significant difference in melting points of clusters (up to 20% for 30 Å clusters on carbon and tungsten substrates [6]). Physically it can be related to the additional energy of the supported cluster due to the contact potential difference [12], the lattice mismatch at the cluster-substrate interface [13], and the cluster-substrate van der Waals interaction [14] that can lead to the shifts of cluster's lattice parameter and melting temperature. Therefore, the understanding of the physical reason for the size shift of cluster's melting point and lattice parameter needs the determination of the influence of the substrate on cluster's properties.

In this paper the theoretical description of the melting point and the lattice parameter shifts for the supported nanoclusters of various metals carried out in the framework of the uniform approach using the vacancy mechanism taking into account the cluster-substrate interaction is proposed. In the framework of this mechanism the change of cluster properties with decrease of its size is described as a result of anharmonic oscillations of cluster atoms, leading to the decrease of the energy of vacancy formation and, hence, to the additional creation of vacancies in cluster with decrease of its size.

The estimations show that the change of the energy of vacancies formation in a cluster with size of ~30 Å due to the van der Waals interaction with substrate can be as much as $\delta\mu_{cluster}^{vdW}\big|_{h=30\text{Å}} \sim 0.2eV$, while the energy change due to the cluster-substrate lattice mismatch, to the contact potential difference at the interface, as well as to the adsorption of atoms at cluster surface [16] under the pressure ($P < 10^{-7}\ torr$) does not exceed $0.05eV$.

Therefore it can be concluded that the major effect on the shift of the melting point and the lattice parameter of supported nanoclusters under the high vacuum conditions ($P < 10^{-7}\ torr$) is produced by the van der Waals cluster-substrate interaction.

This effect can be described considering the van der Waals interaction of a supported metal cluster with bulk metal or semimetal substrate [12,14]. Van der Waals forces are of the electromagnetic character and arise due to the mutual polarization of neutral atoms of the interacting objects. The presence of these forces can lead to the change of the chemical potential of a cluster. In the case the change is positive, it can be interpreted as an additional thermodynamic potential per atom of a cluster, arising due to the interaction with substrate. It results in the decrease of the binding energy of atoms in a lattice and, therefore, in the increase of the probability of vacancies formation in a cluster. Besides van der Waals forces, the formation of vacancies in a cluster occurs due to the presence of cluster surface, that results in the dependence of the energy of vacancy formation on its distance from cluster surface, as well as on cluster shape [17]. Under assumption that the clusters under investigation [1-5] satisfy the requirement $l/h > 1$ (where $l$=0.7-10.0 nm and $h$=0.3-5.0 nm are cluster's lateral size and height, consequently), the cluster surface can be considered as a flat and the influence of the boundary effects related to cluster's shape can be neglected.

The equilibrium concentration of vacancies in a cluster can be found from the requirement of the minimum of the free energy of cluster's vacancies subsystem taking into account the interaction with substrate:

$$\left.\frac{\delta F}{\delta n}\right|_{n=\overline{n}} = 0 \qquad (1)$$

where $\overline{n}$ is the average concentration of vacancies in a cluster. The interaction of vacancies at the distance more than the character size of the vacancy (the possibility of divacancy formation) is attractive, while at the distance less than the characteristic size of the vacancy it has a character of a solid core (it is impossible to locate two vacancies at the same site of the lattice). Assuming the equilibrium concentration of vacancies much less than the atomic density in a cluster, the vacancies subsystem of a cluster can be considered as a van der Waals' gas [18]. In this case the free energy of the vacancies subsystem after decomposition for virial coefficients is given by [19]:

$$F = F_{id} + \frac{N^2 T B(T)}{V}\ ; B(T) = \frac{1}{2}\int \left(1 - \exp\left(-\frac{U_{12}}{T}\right)\right) dV \qquad (2)$$

where $N$ is the number of vacancies in a cluster, $V$ is the cluster volume, $U_{12}$ is the interaction potential of vacancies, and $F_{id}$ is the free energy of non-interacting vacancies system. Considering $U_{12}$ as a rectangular potential well of depth $E_d$ (the energy of divacancy formation), the virial coefficient $B(T)$ is expressed as:

$$B(T) = b + 7b\left(1 - \exp\left(\frac{E_d}{T}\right)\right) \qquad (3)$$

where $b = \frac{2\pi r_0^3}{3}$, $r_0$ is the characteristic size of a vacancy. The free energy $F_{id}$ of non-interacting vacancies system is given by [17]:

$$F_{id} = \int E_v \overline{n} dV + T\int \left[\overline{n}\ln\left(\frac{\overline{n}}{n_0}\right) - \overline{n}\right] dV\ ; E_v = E_v^B - \delta\mu\ ; \qquad (4)$$

$$E_v^B = \frac{1}{h}\int_0^h \left(E_B - (E_b - E_S)\exp\left(-\frac{x}{2a}\right)\right) dx$$

Where $E_B, E_S$ are the energies of vacancy formation in the volume and at the surface of a cluster, consequently, $n_0$ is the atomic density for a cluster, $a$ is the cluster lattice constant, $\delta\mu$ is the additive chemical potential of a cluster arising due to its interaction with substrate [14]. In the expression (4) the energy of the vacancy formation is assumed depending on its distance from cluster surface, and $E_v^B$ is the energy of vacancy formation averaged for a cluster. Substituting the expression (4) into (2), the free energy is obtained:

$$F = \int E_v \overline{n} dV + T\int \left[\overline{n}\ln\left(\frac{\overline{n}}{n_0}\right) - \overline{n}\right] dV - nT\ln(1 - nb) - 7n^2 bT\left(1 - \exp\left(\frac{E_d}{T}\right)\right) \qquad (5)$$

The equation for the equilibrium concentration of vacancies $\overline{n}$ in a cluster follows from (5) and (1) as:

$$E_v^B - \delta\mu + T\ln\frac{\overline{n}}{n_0} - T\ln(1 - \overline{n}b) + \overline{n}T\frac{b}{1-\overline{n}b} - 14\overline{n}bT\left(1 - \exp\left(\frac{E_d}{T}\right)\right) = 0 \qquad (6)$$

Where $\delta\mu$ can be calculated if the complex dielectric constants of substrate and cluster are known [14,12]. The effect of the cluster surface taken into account in the expression for the energy of vacancy formation, the value $\delta\mu$ can be calculated for a film of thickness $h$ deposited onto substrate [14,12]:

$$\delta\mu = \frac{a^3}{z}\frac{\hbar\overline{\omega}}{8\pi^2 h^3}\ ;\ \overline{\omega} = \int_0^\infty \frac{(\varepsilon_f(i\xi)-1)(\varepsilon_f(i\xi)-\varepsilon_s(i\xi))}{(\varepsilon_f(i\xi)+1)(\varepsilon_f(i\xi)+\varepsilon_s(i\xi))} d\xi \qquad (7)$$

Where $i\xi = \omega$ is the imaginary part of the electromagnetic field, $\varepsilon_s, \varepsilon_f$ are the dielectric constants of substrate and cluster materials, consequently, and $z$ is the number of the nearest neighbors in a cluster. Under assumption of the validity of the expression $\varepsilon_j(i\xi) = 1 + \frac{4\pi\sigma_j}{\xi}, j = s, f$ for the dielectric constants $\varepsilon_f, \varepsilon_s$ [12,20], the value $\delta\mu$ is given by:

$$\delta\mu = \frac{a^3}{z}\frac{\hbar(\sigma_f - \sigma_s)}{4\pi h^3}\frac{\sigma_f}{\sigma_s}\ln\left(\frac{\sigma_f + \sigma_s}{\sigma_f}\right) \qquad (8)$$

where $\sigma_f, \sigma_s$ are the conductivity of cluster and substrate, consequently.

The conductivity $\sigma_s$ of a cluster can be calculated by taking into account the contributions to the intrinsic resistivity of the cluster material $\rho_0$ and the additional resistivity $\rho_{vac}$ arising due to the additional scattering of conduction electrons from vacancies. The additional resistivity $\rho_{vac}$ depends on the concentration of vacancies defined by the additional chemical potential arising due to the presence of a substrate. Thus, the expression for the cluster conductivity takes the form:

$$\sigma_f = (\rho_0 + \rho_{vac})^{-1} \, ; \, \rho_{vac} = \rho_{vac}^0 \frac{\bar{n}}{n_0} \tag{9}$$

Where $\rho_{vac}^0$ is the additional resistivity per a vacancy that does not depend on $\bar{n}$. The expressions (6), (8) and (9) allow to calculate the equilibrium concentration of vacancies in a cluster $\bar{n}$ taking into account the cluster-substrate interaction. The theoretical dependence of the relative concentration of vacancies for Au cluster on graphite surface as function of cluster height at room temperature ($T$=300 K) calculated from the equations (6), (8) and (9) is plotted in Fig. 1.

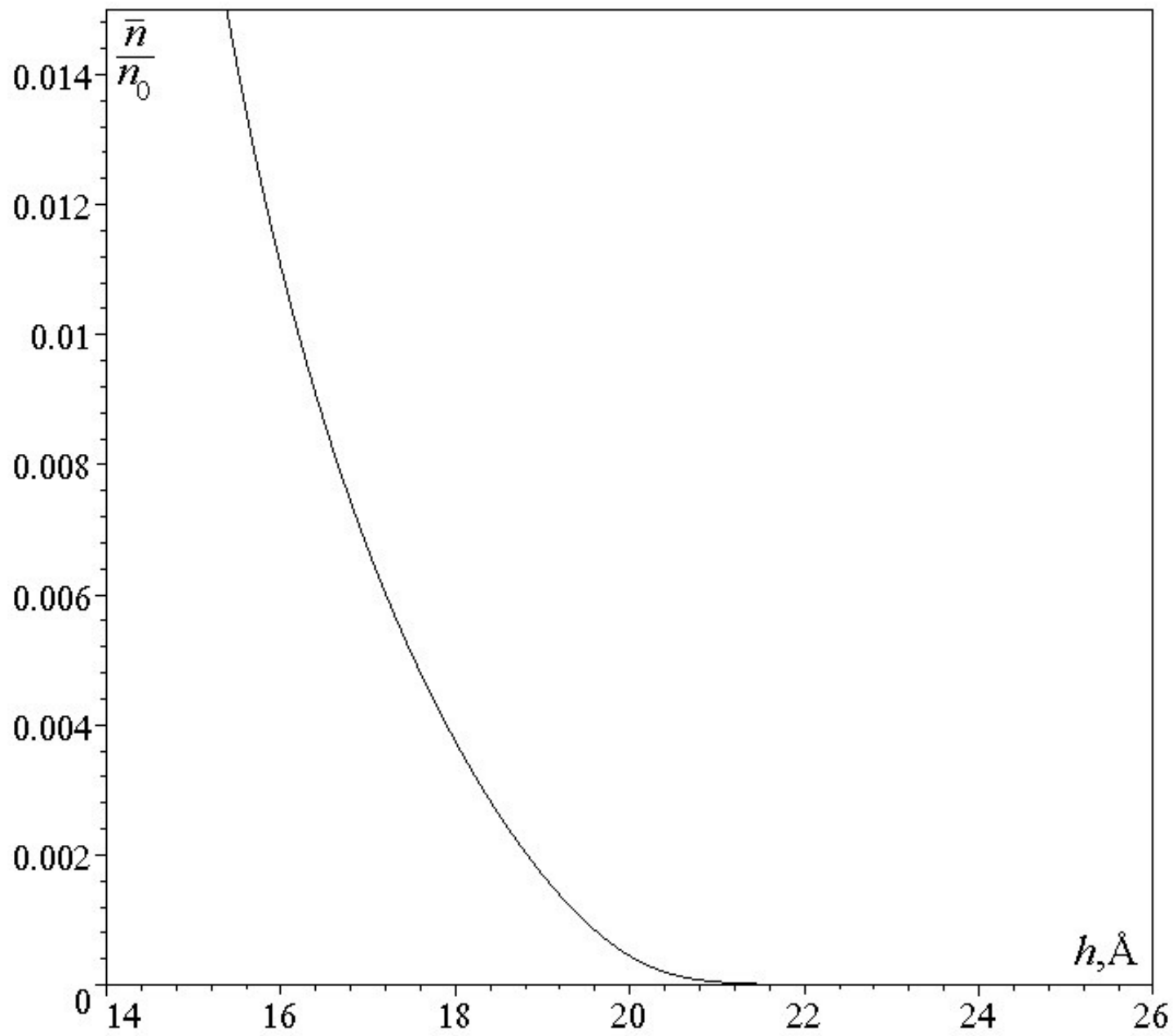

**Fig 1.** The equilibrium concentration of vacancies $\bar{n}$ in Au cluster on graphite at $T$=300 K relative to the atomic density $n_0$ calculated as function of cluster height $h$.

It is seen that the concentration of vacancies sharply increases for cluster height less than 22 Å, that can result in the modification of the modulus of elasticity of cluster and, as a consequence, to the shifts of the lattice parameter and the melting point [18].

The melting point can be calculated taking into account the dependence of the modulus of elasticity on the concentration of vacancies in a cluster [18]. Using the vacancy model of melting [14] and considering vacancies as dilatation centers, it is possible to calculate the shear modulus $K'$ as function of the concentration of vacancies [14]:

$$K' = K - \frac{4}{15}\pi^4 (K\Delta V)^2 \frac{\bar{n}\left(1-\frac{\bar{n}}{n_0}\right)^2}{T\left[1+7\frac{\bar{n}}{n_0}\left(1-\frac{\bar{n}}{n_0}\right)^2\left(1-\exp\left(\frac{E_d}{T}\right)\right)\right]} \tag{10}$$

Here $K$ is the shear modulus of the defect free material and $\Delta V = 0.1 \div 0.2a^3$ is the dilatation volume. Defining the melting point of a cluster from the Born's criterion $K'(T_{melt}) = 0$ [18], the dependence of the melting point on cluster height can be obtained from the equation (10). The dependence of the melting point of Au clusters supported on carbon and tungsten substrates as function of cluster height, calculated from (6), (8), (9), (10), as well as the dependence calculated without the influence of the substrate, are presented in Fig. 2.

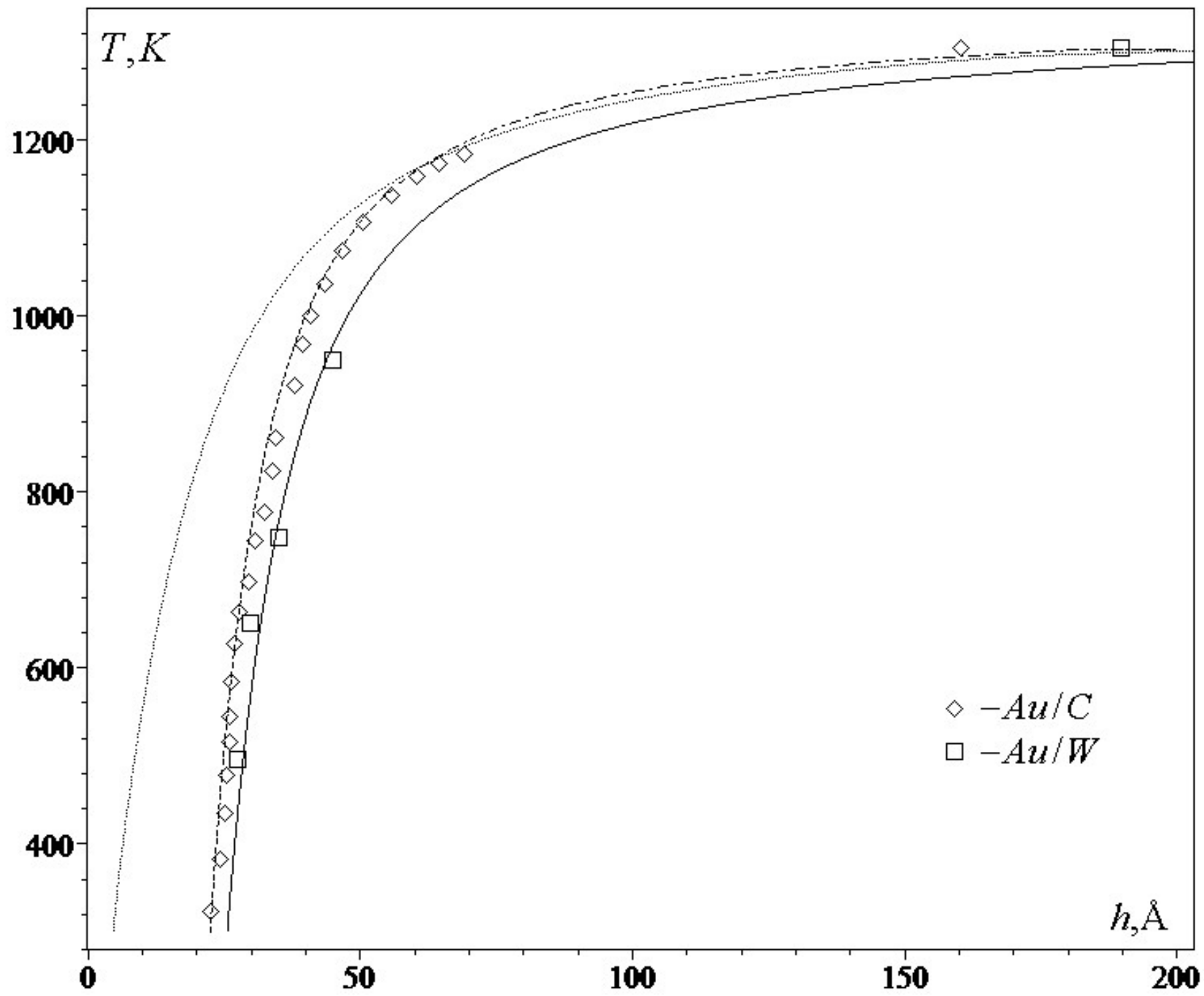

**Fig. 2.** The dependence of the melting point $T_{melt}$ of Au cluster on its heigh $h$. Solid and dashed curves are the theoretical dependences calculated using equations (6), (8), (9) and (10) for Au clusters on W and C, consequently. Dotted line is the dependence calculated without the influence of a substrate. Points are the experimental data [1,6].

The plots in Fig.2 evidence that the substrate plays an important role in the shift of the melting point of clusters. It is seen that different substrates result in the significant shift of the melting point of clusters that is satisfactorily described by the proposed model (see Fig.2).

The formation of extra number of vacancies in a cluster compared to the bulk metal leads to the reduction of the lattice parameter of a cluster that is related to the vacancy concentration [21]:

$$\frac{\Delta a}{a_0} = \frac{1}{3}\frac{V_x}{V_0}\overline{n} \tag{11}$$

Where $a_0$ is the lattice parameter in the absence of vacancies, $\Delta a$ is the change of the lattice parameter due to vacancies, $V_0$ is the atomic volume of a defect free crystal, and $V_x$ is the change of the atomic volume caused by the presence of one vacancy. Using the relationship between $V_0$ and $n_0$ the lattice parameter in the presence of vacancies can be expressed as:

$$\frac{a}{a_0} = 1 - \frac{1}{3}\frac{\overline{n}}{n_0} \tag{12}$$

The expression (12) shows that the additional formation of vacancies in a cluster results in the decrease of the lattice parameter.

Equations (6), (8), (9), (10) and (12) allow to calculate the dependence of the lattice parameter on cluster height. Such dependences for the clusters of Cu, Au and Ag deposited on graphite at room temperature are presented in Fig.3. The parameters used in the calculations of the dependencies plotted in Figs. 1-3 are summarized in Table 1.

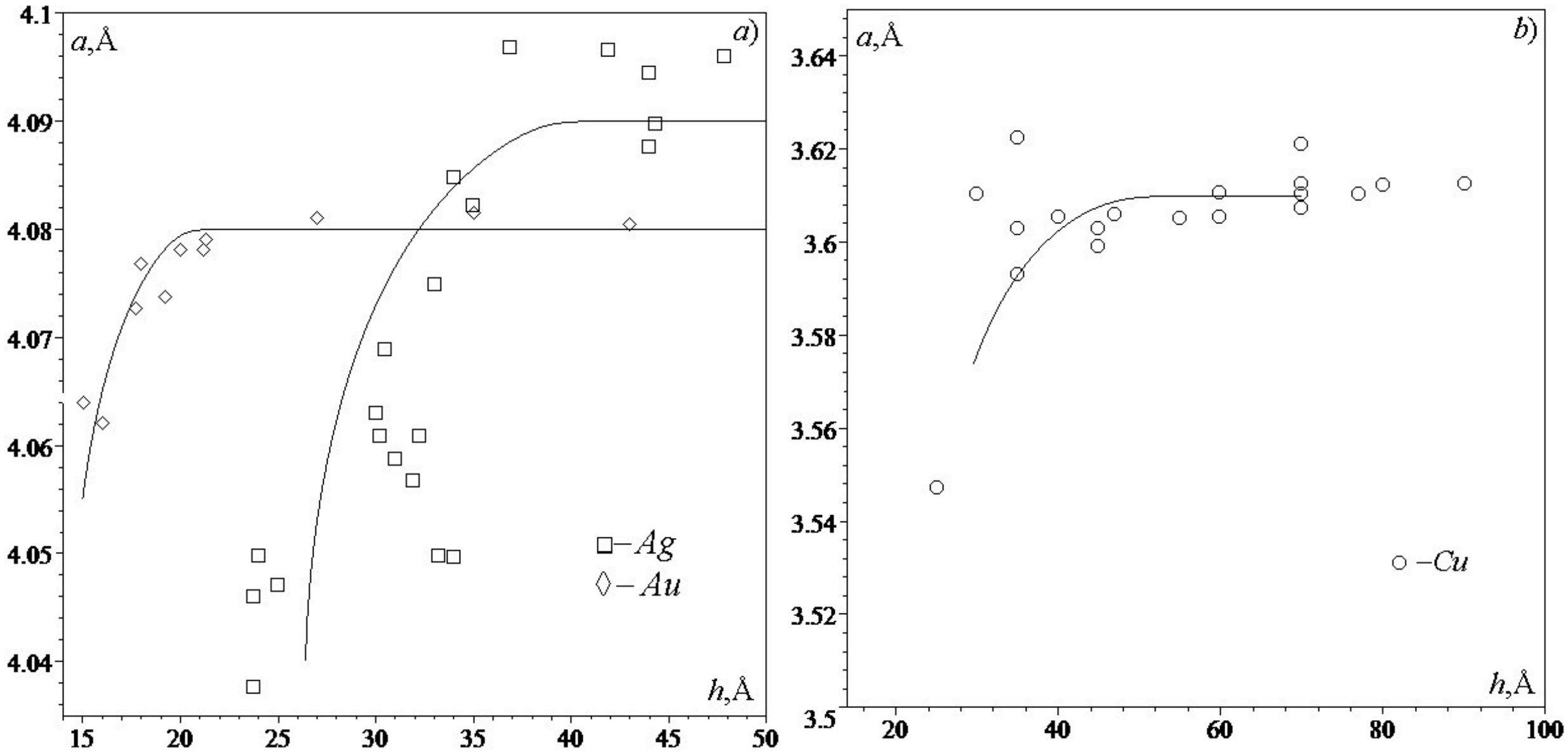

**Fig. 3.** The lattice parameter $a$ for Au, Ag (*a*) and Cu (*b*) clusters on graphite as function of cluster height $h$ at room temperature. Solid lines are the theoretical dependences calculated using equations (6), (8), (9), (10) and (12). Points are the experimental data [2].

**Table 1.** Parameters used in the calculations.

| Parameter | Au | Ag | Cu |
|---|---|---|---|
| The energy of vacancy formation in the volume, eV | 0.9 [23] | 1.11 [15] | 1.28 [15] |
| The energy of vacancy formation at the surface, eV | 0.25 [23] | 0.16 [15] | 0.35 [15] |
| Divacancy formation energy [23], eV | 0.1 | 0.1 | 0.05 |
| Lattice parameter [24], Å | 4.08 | 4.09 | 3.61 |
| Specific conductivity [23], $10^{-6}$ Ω·cm | 2 | 1.5 | 1.55 |

Fig. 3 shows that the additional formation of vacancies in a cluster due to its interaction with substrate results in the decrease of the lattice parameter that agrees with the experimental data.

The additional energy of a cluster per one atom due to its interaction with substrate can be estimated in the frames of several mechanisms taking into account the contact potential difference, the lattice mismatch at the interface, the van der Waals interaction with the substrate, as well as the adsorption and the chemical reactions at cluster surface. The estimations will be carried out for the cluster of 30 Å in height. According to the equations (8) and (9), the additional energy of a cluster per one atom due to the van der Waals interaction with substrate is $\delta\mu^{vdW}_{cluster}\,\big|_{h=30\text{Å}} \sim 0.2eV$.

The shift of the cluster chemical potential due to the presence of the cluster-substrate interface is given by [12]: $\delta\mu = A\exp(-\kappa x)\,; A = \frac{\kappa_2(\mu_2-\mu_1)}{\kappa_1+k_2}$ ; $\kappa^2_{1,2} = 4\pi e^2 \frac{n^0_{1,2}}{kT}$ where $\mu_1, \mu_2$ are the chemical potentials of graphite (1) and Au (2), $n^0_{1,2}$ is the electron density, $\kappa = \frac{1}{L_{sc}}$ is the reverse Debye screening radius.

The additional energy of a cluster obtained due to the contact potential difference can be estimated assuming that the influence of a substrate is considerable only within the first monolayer of atoms at the interface: $\delta\mu_c \sim \frac{1}{V}\int \delta\mu dV \sim \frac{\delta\mu\pi\, r^2 1/\kappa}{\pi r^2 h} \sim \frac{\delta\mu}{h\kappa}$ , where $V, h$ are the volume and the height of a cluster. In this case the value of the additional energy due to the contact potential difference equals to
$\delta\mu^{\varphi}_{cluster}\,\big|_{h=30\text{Å}} \approx -0.05eV$.

The additional energy of a cluster per one atom due to the lattice mismatch at the interface can be estimated considering that the characteristic temperature of the transition to the disproportionate phase is about 500 K. [26]. Thus, the additional energy per one atom due to the this mechanism is $\delta\mu_{cluster}^{lattice} \sim 0.05eV$ .

The maximal additional energy per one atom due to the adsorption and the chemical reactions at the cluster surface can be estimated as $\delta\mu_{cluster}^{ad} = \overline{N}_1 E_{fr}$ , where $\overline{N}_1$ is the number of atoms adsorbed at cluster surface, and $E_{fr}$ is the hemisorption cohesive energy ($E_{fr} \sim 1 \div 1.5eV$). Fractional coverage of the cluster surface is 0.001 under $P = 10^{-9}\ torr$, so the additional energy per one atom of cluster due to the adsorption and the chemical reactions is less than 0.01 eV. Note that this value increases with pressure and under $P = 10^{-5} torr$ it may reach, $\sim 0.7eV$ that may be the reason of the experimentally observed [2,27] lattice extension under the low vacuum conditions.

Summarizing the above estimations, it is seen that under the high vacuum conditions $\delta\mu_{cluster}^{vdW}\Big|_{l=30\text{Å}} \gg \delta\mu_{cluster}^{lattice}, \delta\mu_{cluster}^{ad}, \delta\mu_{cluster}^{\varphi}$. Therefore under these conditions in the framework of the considered mechanism the shift of the melting point and the lattice parameter of clusters with decrease of their size is defined by the van der Waals cluster-substrate interaction.

The work has been partially supported by the Russian Foundation for Basic Research, grants No. 02-00759a and 07-02-01372a.